\documentclass[aps,amssymb,prl,amsmath,twocolumn]{revtex4}
\usepackage{graphics}
\usepackage{psfrag}
\usepackage{subfigure}
\usepackage{epsfig}

\def\bn{\begin{enumerate}}
\def\en{\end{enumerate}}
\def\be{\begin{equation}}
\def\ee{\end{equation}}

\begin{document}
\bibliographystyle{prsty}

\title{Numerical Tests of Constitutive Laws for Dense Granular Flows}
\author{Gregg Lois$^{(1)}$}
\author{Ana\"el Lema\^{\i}tre$^{(1,2)}$}
\author{Jean M. Carlson$^{(1)}$}
\affiliation{
$^{(1)}$ Department of physics, University of California, Santa Barbara, California 93106, U.S.A.}
\affiliation{$^{(2)}$ L.M.D.H. - Universite Paris VI, UMR 7603, 4 place Jussieu - case 86, 75005  Paris - France}
\date{\today}

\begin{abstract}
We numerically and theoretically study macroscopic properties of dense, sheared granular materials.
In this process we first introduce an invariance in Newton's equations, explain how 
it leads to Bagnold's scaling, and discuss how it relates to the dynamics
of granular temperature.  
Next we implement numerical simulations of granular materials
in two different geometries--simple shear and flow down an incline--and
show that measurements can be extrapolated
from one geometry to the other. 
Then we observe non-affine rearrangements of clusters
of grains in response to shear strain and show that fundamental 
observations, 
which served as a basis for the Shear Transformation Zone (STZ) theory of amorphous solids~\cite{falk98,falk00}, can be reproduced  
in granular materials.
Finally we present constitutive equations for granular materials as proposed in~\cite{lemaitre02a}, based on 
the dynamics of granular temperature and STZ theory,
and show that they match remarkably well with our numerical data from both geometries.
\end{abstract}
\maketitle

\section{I. Introduction} 
Historically reserved to the engineering 
community~\cite{SW68,brown70,ROS70,nedderman92},
granular materials recently emerged as a new field of 
study for physicists~\cite{JN92,mehta94,jaeger96,JNB96,duran97,deGennes98,rajchenbach00}:
a state of matter that is not 
classifiable according to the
traditional ternary--solid, liquid or gas--
and requires scientists to rethink the foundations of statistical 
physics and thermodynamics~\cite{Kad99,EG02}.
The first theories of granular materials were motivated 
primarily by the need to predict the creep motion 
of soils and their stability properties.
Inspired by continuum theories of plasticity, 
these approaches  were restricted 
to quasi-static deformation and incipient 
failure~\cite{SW68,brown70,ROS70,nedderman92,HW92,HAR95,HZ96}.
In physics, initial progress came in understanding dilute granular materials
with the development of kinetic theory~\cite{savage79,SJ81,JS83,LSJ+84,JR85,LS87,SAV89,CAM90,HR94,SG95,SGN96,SG98,dufty01}.
But kinetic theory is based on the assumption that interactions between
grains occur during instantaneous binary collisions,
a condition that is realized by very few ``rapid'' flows outside the laboratory.

Traditional approaches thus leave us with little understanding of the elementary
mechanisms of deformation in granular materials.
It is no wonder that the elaboration of physically inspired constitutive equations 
for granular materials is currently facing a collection of controversial issues 
which challenge its most fundamental aspects. 
Common points of contention include:
(i)
the relevant features of the grain-grain 
interaction--{\it e.g.\/} Hertzian versus 
hard-sphere repulsion--~\cite{hertz81,johnson85,rajchenbach00};
(ii)
the domain of applicability of kinetic theories,~\cite{SG95,SG98,azanza98,azanza99,MitaraiN04};
(iii)
the observation and micromechanical origin of Bagnold scaling in dense 
flows~\cite{bagnold54,bagnold66,HZC+02,silbert01,silbert03,rajchenbach04};
(iv)
the plausible need to formulate ``non-local'' 
constitutive equations~\cite{MLT99,MTL00,chevoir01,EH02,midi04}
due to the existence of chain forces~\cite{CS79,DD72,MJN98,HBV99a,HBV99b}.
Uncertainties about these questions have fueled a wealth of theoretical models.
Some models posit that frustrated rotation plays a preeminent role in jamming~\cite{Kan79,MHN02}.
Many models propose extensions of kinetic theory by using either granular temperature~\cite{SAV98},
introducing strongly density-dependent viscosities~\cite{LBL+00,BEL02,BLS+02},
or incorporating a quasi-static stress at the internal friction angle
of the material~\cite{savage83,JJ87,JNJ90,JA91,AJ92,louge03}.
Other models are based on the introduction of chain-forces~\cite{MLT99,MTL00,chevoir01},
activated processes~\cite{PG96}, ``granular eddies''~\cite{EH02}, coexisting 
liquid and solid microphases in a Ginzburg-Landau formulation~\cite{ATV99,AT01,VTA03},
or 'spots' of free-volume associated to cooperative diffusion~\cite{Baz04,Baz05}.

Several of the above-mentioned models conjecture
jamming mechanisms--frustrated rotations, chain forces, granular eddies--which
are so specific to granular materials that they do not allow for connection with 
jamming in other materials.
Our approach rests on the viewpoint that,
since jamming is observed in many amorphous systems,
it is likely that it has a common origin.
It was thus proposed~\cite{lemaitre02a} that a rescaling of the dynamics of 
perfectly hard granular materials would make it possible to map their properties onto 
more typical glass formers.
Building on the analogy between granular materials and metallic glasses then 
provides a local mechanism of jamming
on the basis of the Shear Transformation Zone (STZ) theory of plasticity~\cite{falk98,falk00}.

The goal of the present paper is both to discuss some of the microscopic assumptions 
underlying the construction of constitutive equations for granular materials and 
test the specific predictions of the STZ theory formulation 
of such equations~\cite{lemaitre02a}.
Our approach is based on the following expectations concerning the 
four points of contention identified above:
(i) major properties of dense granular flows can 
be captured by perfectly hard grains; (ii) jamming involves variations 
of the frictional stress in regimes where the collisional contributions-- those predicted by kinetic theory-- are negligible;
(iii) for flows of perfectly hard grains, Bagnold's scaling is relevant at all densities and arises 
from an invariance in the equations of motion; and
(iv) the rheology of dense granular materials is local in a sense to be defined further.

Indeed, (i) a recent work by Campbell has shown--convincingly to us--that 
most natural and experimental flows occur in a regime of strain rates where grains can 
be considered as hard-spheres~\cite{CAM02}. Although this question may still be debated, 
we will take this for granted and restrict our study to a material composed
of perfectly hard grains.
(ii) The breakdown of kinetic theory in dense flows has now been 
characterized in several numerical and 
experimental studies~\cite{ZC92,CAM93,NHT95,CCH95,PC96,CKR+04}
and indicates that jamming is associated with changes in the contribution of long-lasting 
contacts to the stress tensor, not the collisional contribution.
We will take this idea for granted in the present work, but we do believe
that a quantitative analysis of this statement is needed and will devote a future
paper to this question.
(iii) We will show, following~\cite{lemaitre02a} that in the case of perfectly
hard grains, Bagnold's scaling holds for any density.
Assumption (iv) is strongly inspired by the works by Aranson, Tsimring and coworkers,
who have shown that a Ginzburg-Landau formulation of a fluid-solid mixture~\cite{ATV99,AT01,VTA03}
could account for important properties of granular flows. This suggests that
a local formulation of granular rheology, coupled to hydrodynamic equations, 
can be sufficient to account for a large part of the phenomena 
observed in dense granular flows. 
However, no numerical study has been specifically devoted to this question.

Therefore, we will first study whether dense granular flows
can be described as a local phenomenon, governed by local hydrodynamic equations, 
or whether granular flows must be treated non-locally, relying on the emergence of long-range 
correlations through mechanisms such as force chains.
For this purpose we have implemented Contact Dynamics simulations of 
sheared granular materials
in two different configurations: 
(i) simple shear in a periodic cell and
(ii) thick flows of granular material down an incline plane.
We will show that the measurements obtained in either configuration can be extrapolated
to the other.


Next we follow Falk and Langer~\cite{falk98,falk00} and test for the consistency
of the STZ picture of material deformation advocated by these authors.
Our observation provides further grounding for the analogy between granular
materials and common glass-formers and direct support for the relevance
of STZ theory to granular materials.
A review of the STZ constitutive equations for granular flows will follow and
finally a fit of the theory with our numerical data.

The organization of our paper is as follows.
In Section~II we present equations of motion in the mathematical limit of perfectly hard grains.
We discuss the invariance properties of the equations of motion and how these properties are related to Bagnold's scaling.  We also discuss why the results
for perfectly hard grains are applicable to natural and experimental granular materials.
In Section~III we construct our numerical test, provide algorithmic details on the Contact Dynamics method, 
and compare the rheology of a dense granular materials in a periodic
cell and down an incline plane.
Finally, in Section~IV, we present Falk and Langer's STZ theory, show how it adapts
to granular materials, and conclude with fits of our data.

\section{II. Fundamental results for Perfectly Hard Grains} 

When grains are dry--so that no water bridges induce attraction--and 
of size larger than the micrometer scale--so that no electrostatic 
interaction intervenes--their interaction is purely repulsive.
The interaction results from the elastic deformation of grains 
at contact and the dissipation of energy via friction and collisions.
The complexity of this interaction motivates our first question:
which properties of the grain-grain interaction contribute 
to any particular macroscopic observation?
In some instances, details of the grain-grain interaction 
seem critical: for example, the Hertzian repulsion~\cite{hertz81} is essential 
to understand the acoustic properties of granular materials~\cite{JCV99,MGJ+04}.
Numerical implementations of granular materials have thus relied 
on more or less elaborate models of the grain-grain 
interaction~\cite{CB85,HL98}.

Here we are concerned with dense flows of granular materials and, in particular,
flows down inclines as found in the experiments by Pouliquen~\cite{Pou99}
and numerics by Silbert and coworkers~\cite{silbert01,silbert03}.
For these dense granular flows a recent study by Campbell helps us
assess the importance of the elastic (soft) part of the 
repulsive potential~\cite{CAM02}
versus the limit in which grains appear as perfectly hard spheres.
Campbell presented a detailed analysis of the
different flow regimes obtained in a three dimensional simple shear simulation
of dense granular flows when varying the stiffness $k$ of the repulsion, 
the shear rate $\dot\gamma$, and the mass density $\phi$.
He found that the dimensionless parameter 
$\Upsilon \equiv \frac{k}{\phi D^3 \dot\gamma^2}$, 
where $D$ is the grain size, dictates the character of the flow.
This quantity is directly related to a Mach number which involves the ratio
of the shear velocity $D\,\dot\gamma$ over the sound speed $c_s$: 
$M=D\,\dot\gamma/c_s=1/\sqrt{\Upsilon}$.
The hard-sphere limit corresponds to the situation where sound waves
travel very fast compared to the motion induced by the shear
flow.  This is the limit of very small Mach number, or small
shear rates. Specifically, in the numerics by Campbell,
this limit is reached for Mach numbers below $\sim10^{-2}$.
Since the sound speed in granular materials is of order 100m/s, if
we assume a grain size of order 1mm, the Mach number
is expressible as $M=10^{-5}\,\dot\gamma$.
Therefore, in order to be in the limit where grains behave as if they are perfectly stiff,
it suffices to restrain oneself to shear rates below 1000s$^{-1}$.
Most experimental and natural
situations occur at shear rates far below this limiting value and we can conclude that most flows of granulates are in the limit 
where the soft part of the repulsion 
is entirely masked by the steric exclusion.  To study these flows
it is sufficient to consider properties of perfectly hard grains.
In the following sections we explore how the mathematical limit of 
perfectly hard spheres gives insight into fundamental processes 
which relate to most all experimental shear flows.

\subsection{Equations of motion and Hard Sphere conditions}

The motion of $N$ spherical grains in a $d$-dimensional granular material 
is determined by Newton's equations for the positions $q_i$, 
angular orientations $\theta_i$, momenta $p_i$ and angular velocities $\omega_i$:
\begin{equation}
\frac{dq_i}{dt}=\frac{p_i}{m_i}, \quad
\frac{dp_i}{dt}=\sum_j F_{ij}+ F^{ext},
\label{Newtonseqns}
\end{equation}
\begin{equation}
\frac{d\theta_i}{dt}=\omega_i,\quad
\frac{d\omega_i}{dt}=\frac{1}{I_i} \sum_j R_i \hat{n}_{ij} \times F_{ij}, 
\label{rotNewtonseqns}
\end{equation}
where $F^{ext}$ represents an external force such as gravity, 
$F_{ij}$ represents a contact force on grain $i$ by grain $j$, 
$\hat{n}_{ij}$ is the unit normal vector pointing from grain $i$ to grain $j$, 
$R_i$ is the radius of grain $i$, and $I_i$ is the moment of inertia.

These equations must be complemented with a prescription for the contact forces.  
In the hard sphere limit these contact forces are determined self-consistently
by the conditions that (i) there is no penetration between grain--a force 
is instantaneously created upon contact to impede penetration and 
remains non-zero until the contact is broken--and (ii) by the friction law which 
couples to rotational degrees of freedom.

Important properties of granular materials arise directly from an invariance 
of the equations of motion~(\ref{Newtonseqns},~\ref{rotNewtonseqns}). We now spend some time studying these properties and 
assessing their consequences for macroscopic observations, in particular Bagnold's scaling.

\subsection{Bagnold's scaling}

The success of kinetic theory came in a large part from its ability
to account for the scaling between stress $\sigma$ and strain rate $\dot\gamma$ 
($\sigma\sim\dot\gamma^2$) first observed by Bagnold in dense granular materials~\cite{bagnold54}.
In Bagnold's experiment, the strain rate was set and the stress measured.
Bagnold justified this behavior 
by assuming that the stress measured in his experiments 
resulted solely from binary collisions: the frequency of collisions and momentum
change per collision are each proportional to the shear rate, therefore the stress is 
proportional to the square of the shear rate.

Recent observations have complicated Bagnold's original assessment. On the one hand, 
the observations by Bagnold have been criticized: they may have arisen from 
a secondary instability of the granular flow in his shear cell~\cite{HZC+02}.
On the other hand, Bagnold's scaling has been directly observed by measuring shear stress and strain rate profiles
in numerical simulations
of granular flows down inclines~\cite{silbert01,silbert03},  
and is found to be consistent with experimental observations of the average flow rate 
in the same geometry~\cite{Pou99}.

Although kinetic theory predicts an exponent which agrees with the power law dependence of Bagnold's scaling,
the derivation relies on the assumption that all interactions between grains occur through binary collisions.
This assumption is applicable to dilute flows but is not upheld
for dense flows, such as those originally studied by Bagnold.  A new argument must be formulated for the dense flow regime.

We contend that in both the dense and ``rapid'' flow regimes, Bagnold's scaling arises from
a very fundamental invariance of Newton's equations in the hard sphere limit.
This invariance does not require any of the assumptions of kinetic theory to hold.
Namely, for a granular material free from external forces, the time evolution 
will obey equations~(\ref{Newtonseqns}, \ref{rotNewtonseqns}) with $F^{ext}=0$.
If we now rescale the contact forces by a scalar value $F_{ij} \rightarrow F_{ij} /A$ 
and simultaneously rescale the time $t\rightarrow t \sqrt{A}$, then Newton's 
equations are transformed to read:   
\begin{equation}
\frac{dq_i}{dt}=\frac{p^{new}_i}{m_i}, \quad
\frac{dp^{new}_i}{dt}=\sum_j F_{ij},
\end{equation}
\begin{equation}
\frac{d\theta_i}{dt}=\omega^{new}_i,\quad
\frac{d\omega^{new}_i}{dt}=\frac{1}{I_i} \sum_j R_i \hat{n}_{ij} \times F_{ij},
\end{equation}
where $p^{new}_i = p_i/\sqrt{A}$ and $\omega^{new}_i = \omega_i/\sqrt{A}$.  
This form for Newton's equations is identical to~(\ref{Newtonseqns}) 
and~(\ref{rotNewtonseqns}) with new values for the momenta and angular velocities.  

Under the rescaling of contact forces and time, the positions 
and angular orientations remain unchanged, while the velocities are changed in accordance with the time rescaling.  
If we were to watch a movie of one granular flow where the grains have initial 
velocities $p_i$, $\omega_i$ and watch another movie at half the speed where the 
initial velocities are doubled 
$p_i \rightarrow 2 p_i$, $\omega_i \rightarrow 2 \omega_i$, 
the two movies would look exactly the same in the hard sphere limit.
The difference in the dynamics is that the contact forces measured 
in the second movie would be $4$ times larger than those 
in the first.

This invariance is a property of
{\it perfectly hard grains\/} which must hold
in the inertial regime.  This includes the regime infinitely close to jamming, 
where multi-body interactions dominate collisional terms and
the basic assumptions of kinetic theory fail.
In an experiment, this scaling breaks down only when it is no longer
appropriate to model the experimental system by perfectly hard grains.
Relying on the arguments of Campbell introduced earlier~\cite{CAM02}, we can conclude that most experimental
granular flows are in the regime where it is appropriate to model the
system by hard grains.

This invariance also holds for any value of the restitution
and friction coefficients. The separation of time scales for dissipation
and shear is not a necessary condition for the invariance to be upheld.
The reason is that dissipation occurs at contacts between grains:
dissipative processes are correlated to the motion of grains,
hence the rate of dissipative processes scales with $\dot\gamma$.

\subsection{Granular Temperature}
Because the positions and angular orientations of grains are invariant to a simultaneous change 
in time and force scales, the path that a granular material takes in configuration 
space is also invariant: only the speed along this path is altered.
Therefore, the path that a granular material takes in configuration space 
can be separated from the rate at which events occur along that path.

This observation leads to a natural definition of granular temperature $T$:
the rate at which microscopic events occur on the configuration 
space path is defined as $\sqrt{T}/\langle R \rangle$ (the average grain size $\langle R \rangle$ is inserted
so that $T$ has units of velocity squared).  These microscopic events are not
necessarily collisions, but can also be understood as force fluctuations or minute
displacement of grains allowing for propagation of ``hard-sphere'' noise.

The invariance is directly related to the structure of Newton's equations
and the existence of inertial terms--or accordingly, to the fact that kinetic
energy, no matter how small, is well defined.
The granular temperature defined above as a ``velocity'' along
phase space trajectories is thus naturally proportional to kinetic energy:
\begin{equation}
T_k = \frac{\langle m \rangle}{2} (\langle v^2 \rangle - \langle v \rangle ^2) + \frac{\langle I \rangle}{2}(\langle \omega^2 \rangle - \langle \omega \rangle^2),
\label{kinetictemperature}
\end{equation}
where $m$ is the mass, $v$ the velocity, $I$ the moment of inertia, $\omega$ the angular velocity, and 
brackets denote an average over grains. 
Since the granular and kinetic temperatures are defined up to a constant factor, we will equate them.
This means that at any time, the kinetic energy provides an estimate of the frequency
of elementary events in a multicontact system.

\subsection{What quasi-static limit?}

Consider a uniform granular material in simple shear flow.
If we imagine simultaneously changing the force and time scales then the invariance in 
Newton's equations guarantees that the dynamics will remain unaltered, resulting only in new
velocities that are determined from the time rescaling.  However, under the rescaling, quantities such as the pressure
and strain rate will take different values.  The quantities of interest are those that are 
invariant to the transformation.  These quantities must be formed of ratios of force scales, ratios of time scales,
or ratios of force to time scales.

Denote $p$ as the pressure, $\sigma$ the shear stress, $\dot\gamma$
the strain rate, $\phi$ the mass density, and $D$ the average diameter of grains.
Some important invariant (and dimensionless) quantities are: $\frac{p}{\phi\,D^2\,\dot\gamma^2}$, 
$\frac{\sigma}{\phi\,D^2\,\dot\gamma^2}$, $\frac{p\,D^{d}}{T}$, 
$\frac{\sigma \, D^{d} }{T}$, $\frac{T}{\phi \, D^{d+2}\dot\gamma^2}$, and
$\frac{\sigma}{p}$.  These must be single valued functions 
of {\it density\/} only (independent 
of the strain rate).  This observation, established by the invariance in Newton's equations which holds
for granular flows in the dense and collisional regimes, automatically predicts Bagnold's scaling:  since $\frac{\sigma}{\phi\,D^3\,\dot\gamma^2}$
is a function of density only, it follows that $\sigma \propto \dot\gamma^2$.

This leads to a simple yet remarkable property:
{\it at a given density\/}, changing the strain rate does not lead the system 
toward any different regime or ``phase''. The material does not go closer to jamming because 
the strain rate is scaled down.  The system is always exploring the same trajectories
in phase space, but at a slower speed.  
In short, there is no quasi-static limit for perfectly hard grains.
This is apparent in the jamming phase diagram of granular materials~\cite{O'HernSLN03} where changing
the density allows the granular material to jam, while changing the strain rate does not. 
This property is also directly related to the observation by Campbell
that ``there is no path between inertial (rapid) flow and quasi-static flow 
by varying the shear rate at fixed concentration''~\cite{CAM02}. This is no surprise once
we understand the invariance of Newton's equations for perfectly hard grains:
only two regimes are accessible, jammed or inertial, and density is the only parameter
which controls the statistical properties of a flow of perfectly hard grains.

\section{III. Construction of a Numerical Test}

In constructing a numerical test 
our goals are to measure stress-strain relations when granular temperature and stresses 
are appropriately scaled, show that they compare well
with the standard response of yield stress liquids, and
show that the local rheology measured in simple shear flow matches  
the bulk rheology of a granular flow down an incline.

In order to address these issues, we implement numerical 
simulations of granular materials in two different geometries:
\begin{itemize}
\item We implement simple shear flow in a cell with Lees-Edwards (LE) boundary 
conditions. In this configuration, the density and the shear rate is prescribed and 
the simulation cell is, by construction, translationally invariant.
This grants direct access to averaged quantities of the granular temperature and 
stress tensor. Using this configuration we can characterize the 
steady state relation between stresses, granular temperature and strain-rate and extract
numerically the parameters of a constitutive law for granular materials.  A screenshot of 
this shearing geometry is shown in Figure~\ref{lepic}.

\begin{figure}
\resizebox{!}{.38\textwidth}{{\includegraphics{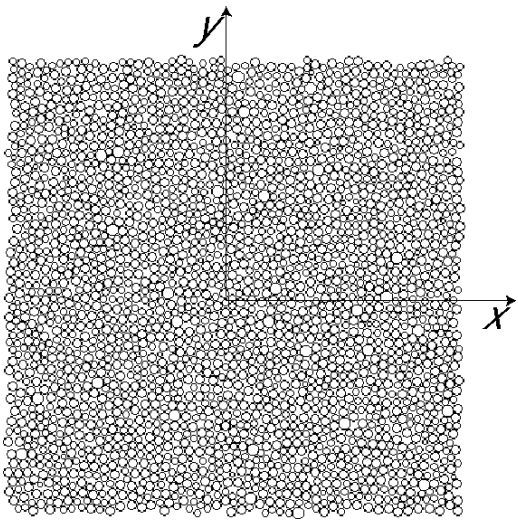}}}
\caption{\label{lepic} Snapshot of a granular material simulation in the simple shear configuration.  
  Each grain has an average velocity in the x-direction given by $\dot\gamma y$, where $\dot \gamma$
is the strain rate.  The center of the cell is defined as $x=y=0$.} 
\end{figure}

\item We implement granular flow down an inclined plane 
made of stationary grains.
The simulation cell is periodic in the direction ($x$) parallel to the plane and
the flow is inhomogeneous in the perpendicular ($y$) direction.
In this configuration, the stresses are prescribed by the angle of the incline.
We perform $x$-averaged, $y$-dependent measurements of granular temperature, 
velocity profiles and strain-rate. 
Large heights of the granular layer grant access to the bulk rheology 
of the flow.  This permits us  
to check the existence of a well-define bulk rheology in the large height
limit, and to compare it with the measurements in simple shear.  A picture of this
shearing geometry is shown in Figure~\ref{toruspic}.

\begin{figure}
\resizebox{!}{.38\textwidth}{{\includegraphics{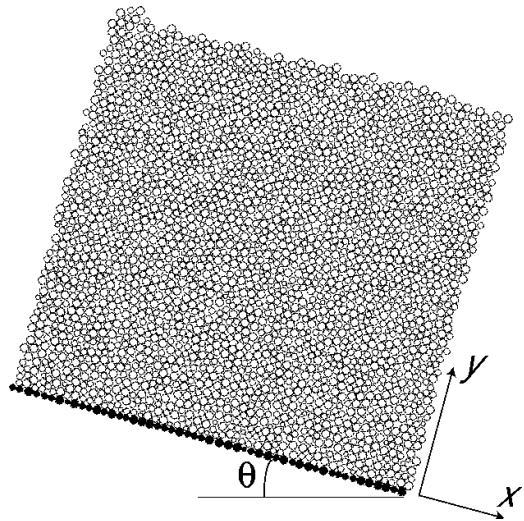}}}
\caption{\label{toruspic} Snapshot of a granular material simulation in the incline flow configuration.  Fixed grains (indicated by filled circles) create a stationary incline at angle $\theta$ on which the flowing grains are accumulated and allowed to flow.  Gravity drives the motion and is directed vertically downward.}  
\end{figure}

\end{itemize}

\begin{figure}
\resizebox{!}{.38\textwidth}{{\includegraphics{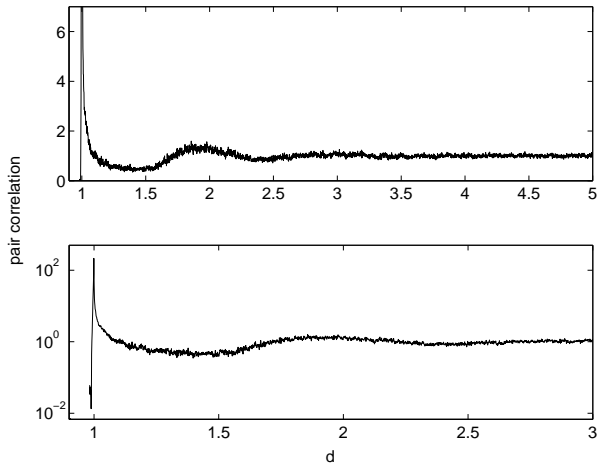}}}
\caption{\label{paircorr}  The pair correlation function for a frictionless granular material at high density in linear (top) and log (bottom) scales.  This function is representative of the pair correlation functions for other densities and for granular materials with friction between grains.  There is a large peak corresponding to contacting grains ($d=1$) and some variation between $d=1$ and $d=3$.  For $d>3$ there is no correlation between grains.  This implies that the material is amorphous.   }
\end{figure}

In order to make a quantitative comparison of the two simulations, we use the 
same material: a two-dimensional polydisperse mixture of constant density grains with the radii drawn 
from a flat distribution with average radius $\langle R \rangle$ 
and width $\sigma$.  
For all of the simulations in this paper we set $\sigma/\langle R \rangle = 0.5$, using $\langle R \rangle = 0.7$.
This distribution prevents crystallization and produces 
an amorphous granular material, as can be seen from measurement of the 
pair correlation function in Figure~\ref{paircorr}.  The horizontal axis ($d$) in this figure
corresponds to the distance between a pair of grains, divided by the sum of their radii.  This normalizes the 
figure so that $d=1$ corresponds to contacting grains.  
Other than this peak, the function has some small variation (which implies a correlation) between $d=1$ and $d=3$.  However, there
is no correlation beyond $d=3$.  Because there is no large scale correlation, this implies that the granular material is amorphous.  

In this paper we present data for normal and tangential
coefficients of restitution given by $e_n=e_t=0$.
It was shown by Chevoir et al~\cite{chevoir01b} that the regimes
reached by granular materials are almost independent of the restitution coefficients below
some threshold around $e_n  = 0.7$. Different friction coefficient have been
used: frictionless grains and $\mu=0.4$.

Our simulations of granular materials rely on the Contact Dynamics
algorithm~\cite{moreau88,jean92,moreau93,MOR94,moreau96}.
We refer to the literature for technical details and present below a brief overview of 
the method, supplemented by a single addition we made in order to construct 
a Lees-Edwards simulation cell for frictional granular materials.

\subsection{Contact Dynamics}

A Contact Dynamics algorithm was constructed to carry out numerical simulation of 
spheres interacting through the enforcement of hard sphere conditions~\cite{moreau88}.

When a contact occurs there is an non-continuous force created that 
prevents the contacting grains from penetrating.  
The magnitude of this force is chosen to ensure that the final 
relative velocity $u$ of the grains is related to the initial 
relative velocity $u^0$ via the equations
\begin{equation}
u_n = -e_n u^0_n,
\label{restitution}
\quad
u_t = e_t u^0_t,
\end{equation}
where $e_n$, $e_t$ are constant restitution coefficients that will depend on 
the shape and consistency of the grains, and the $n$, $t$ subscripts represent
the normal and tangential parts of the relative velocity with respect to the contact.  
At each time step, the algorithm computes the contact forces by first ensuring 
that these relations hold at each contact. 
To include friction, the Contact Dynamics algorithm ensures that the resulting 
tangential force $F_t$ is less than or equal to $\mu F_n$ where $\mu$ is the 
friction coefficient between grains and $F_n$ is the normal force.  
If this constraint does not hold, then the algorithm sets $F_t=\mu F_n$ 
in order to comply with Coulomb friction.  

In this way the Contact Dynamics algorithm calculates, at each time step, 
contact forces that are consistent with Newton's equations and the hard sphere contact law.

\subsection{SLLOD equations for simple shear flow}

Lees-Edwards (LE) boundary conditions permit us to prescribe the deformation of 
a material by controlling the positions of the image cells~\cite{LE72}.
In all of the simulations presented here, we impose a constant 
strain rate $\dot{\gamma}$ so that a grain at position $y$ has
an average velocity of $\dot\gamma y$ in the $x$-direction (see Figure~\ref{lepic}).  

It was recognized in early implementations of LE boundary conditions
that when deformation is applied through the image cells, 
the information needs time to propagate from the cell boundaries to its center. 
In order to ensure rapid propagation of this information and 
prevent the boundaries between cells from making unphysical contributions to the motion,
it is necessary 
to modify Newton's equations by introducing so-called SLLOD terms. 
These terms can be understood as a sort of ``shear bath'', 
with all particles in the cell being directly coupled to the overall 
deformation~\cite{EM84}.   
In practice, the SLLOD terms introduce a mechanical perturbation to the equations of motion 
that gives each grain an average velocity consistent with simple shear flow.
If we separate the momentum $p_i$ of each grain $i$ into the average part $m_i \dot\gamma y_i$ 
and fluctuating part $\tilde{p}_i$, so that $p_i=m \dot\gamma y_i + \tilde{p}_i$,  
then the SLLOD equations read:
\begin{equation}
\frac{dq_i}{dt}=\frac{\tilde{p}_i}{m_i}+ \hat{x} \dot{\gamma} (q_{i} \cdot \hat{y}),
\quad
\frac{d\tilde{p}_i}{dt}=\sum_j F_{ij}- \hat{x} \dot{\gamma} (\tilde{p}_{i} \cdot \hat{y}).
\label{SLLODpos}
\end{equation}
The equation for the position $q_i$ is simply the result of writing the momentum
in terms of an average and fluctuating part.  The equation for $\tilde{p}_i$ contains
a new term $\hat{x} \dot{\gamma} (\tilde{p}_{i} \cdot \hat{y})$ which forces the shear flow.
Since every grain in the primitive cell is acted upon by this mechanical force, 
the constant strain rate is imposed on all of the grains simultaneously at the 
beginning of the simulation.  Furthermore it can be proven that, in the LE 
geometry, the SLLOD equations give an exact representation of simple shear flow 
arbitrarily far from equilibrium~\cite{EM84,EM90}.

For a granular material with non-zero friction coefficient $\mu$,
the equations of motions should incorporate rotations of the grains (for $\mu=0$ the tangential contact force is
always zero and there is no rotation). 
It is expected that a SLLOD term should arise in the equations of motion 
for the angular velocity since, in the linear velocity profile 
indicative of simple shear flow, the top and bottom of every grain 
should be moving with slightly different velocities.  
This will give each grain an average rotation of $\dot{\gamma}/2$ which must 
be incorporated in equation~(\ref{rotNewtonseqns}) just
as the average velocity $\hat{x} \dot{\gamma} (q_{i} \cdot \hat{y})$ 
was incorporated in equation~(\ref{SLLODpos}).
This leads to the following equations:
\begin{equation}
\frac{d\theta_i}{dt}=\tilde{\omega}_i+\frac{\dot{\gamma}}{2} ,
\quad
\frac{d\tilde{\omega}_i}{dt}=\frac{2}{m_i r_i^2} \sum_j R_i \hat{n}_{ij} \times F_{ij},  
\label{SLLODrot}
\end{equation}
where $\tilde{\omega}_i$ denotes the fluctuating part of the angular velocity 
and we have inserted the moment of inertia of constant density disks in two dimensions.  

Equations~(\ref{SLLODpos}) and (\ref{SLLODrot}) now give an exact 
representation of simple shear flow for a frictional granular material 
arbitrarily far from equilibrium.  

The primary interest of this procedure is that it permits us to simulate
a sheared granular material with a homogeneous shear rate.
Experimental procedures, {\it e.g.\/} in a Couette cell, do not guarantee
that the strain rate is homogeneous: the existence of walls induce
a non-uniformity of the flow and possibly localization of the deformation.
Our protocol grants direct access to the rheology of the granular
material in a self-averaging situation.

\subsection{Macroscopic Quantities}
We define the stress tensor $\Sigma^{\alpha \beta}$ via Cauchy's equation:
\begin{equation}
\phi \frac{d}{dt} \langle v^{\alpha} \rangle + \phi \langle v^{\beta} \rangle \partial_{\beta} \langle v^{\alpha} \rangle  
= -\partial_{\beta} \Sigma^{\alpha \beta} + f^{\alpha}_{ext},  
\label{cauchy}
\end{equation}
where $\langle v^{\alpha} \rangle$ is the average velocity in the ${\alpha}$ 
direction (averaged over all grains), $\phi$ is the mass density, and $f_{ext}$ is the external force per volume.  This equation simply
states that the total time derivative of the average velocity (left hand side) is proportional to the divergence of the stress tensor, plus any external forces.
Cauchy's equation gives a definition of the stress tensor from which there exists a procedure to derive the functional form 
of the stress tensor.  This is called the Irving-Kirkwood 
derivation~\cite{EM90} and it yields
\begin{equation}
\Sigma^{\alpha \beta} V = \sum_{i} m_i \tilde{v}_i^{\alpha} \tilde{v}_i^{\beta} + \sum_{i>j} (R_i+R_j) \hat{n}^{{\alpha}}_{ij}F^{{\beta}}_{ij},
\label{thestresstensor}
\end{equation}
where $\tilde{v}_i$ is the fluctuating velocity of grain $i$ determined by $\tilde{v}_i^\alpha = v_i^\alpha - \langle v_i^\alpha \rangle$, and V is the volume of the granular material (or area in two dimensions).  
This symmetric stress tensor can be written in terms of three variables:  the pressure $p$, shear stress $\sigma$, and first normal stress difference $\mathit{N}_1$ defined as
\begin{equation}
\Sigma^{\alpha \beta} =  \left( \begin{array}{cc} p\,(1+\mathit{N}_1) & -\sigma \\ -\sigma & p\,(1-\mathit{N}_1) \end{array} \right),
\label{stressbasis}
\end{equation}
where the signs are chosen so that shear stress and pressure are positive 
in our conventions.

The granular temperature $T$ is measured as
\begin{equation}
T = \sum_{i} v_i^2 - \left(\sum_{i} v_i \right)^2 + \frac{1}{2} \sum_{i} R_i^2 \omega_i^2- \frac{1}{2} \left( \sum_{i} R_i \omega_i \right) ^2
\label{defoftemperature}
\end{equation}
which is proportional to the kinetic temperature 
from equation~(\ref{kinetictemperature}), 
with the factor of $1/2$ coming from the moment of inertia calculated 
for constant density discs.

Lastly, for all of the numerical data that will be presented, we quantify the density of the system by its packing fraction $\nu$.
The packing fraction is defined as the area occupied by grains divided by the total area of the system.  In our simulations, packing fraction is proportional to both the mass density $\phi$ ($\pi \, \phi = 4 \nu \,$) and the number density $n$ ($n \pi \langle R^2 \rangle=\nu \,$).

\section{IV. Test of Local rheology}

We first study the average relation between pressure $p$,  
shear stress $\sigma$, shear rate $\dot{\gamma}$, and granular temperature $T$
in simple shear, using the periodic and translationally invariant LE cell.
Next we compare these results with data obtained for the granular flow 
down an inclined plane.

\subsection{Simple Shear}

\subsubsection{Preliminary test}
\begin{figure}
\psfrag{tl}{\LARGE{$\mathbf{p \ \dot{\gamma}^{-2}}$}}
\psfrag{bl}{\LARGE{$\mathbf{p \ \dot\gamma^{-2}}$}}
\psfrag{tc}{\LARGE{$\mathbf{\dot\gamma = 10^{-2}}$}}
\psfrag{bc}{\LARGE{$\mathbf{\dot\gamma = 10^{2}}$}}
\resizebox{!}{.38\textwidth}{{\includegraphics{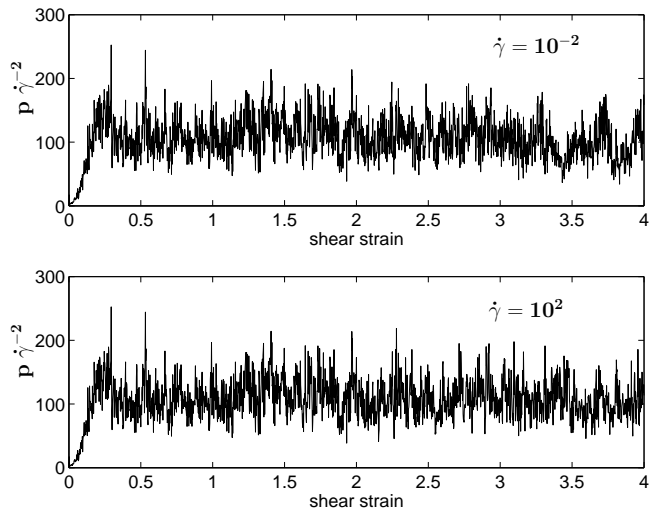}}}
\caption{\label{scaledledata} Raw data of the normalized pressure as a function of total shear for two frictionless granular material with packing fraction 0.8.  Data from simulations with different shear rates $\dot\gamma$ are shown.  The top plot corresponds to $\dot\gamma = 10^{-2}$ and the bottom to $\dot\gamma = 10^2$.   The pressure is normalized by $\dot{\gamma}^2$ which collapses the two data sets on to one master curve (i.e. with this rescaling the top and bottom traces appear essentially identical, up to numerical noise), as predicted by the invariance for hard sphere systems.}
\end{figure}

In all of the simple shear simulations presented here we have simulated $2500$ grains
in a square primitive cell, although we have conducted a limited 
number of simulations with up to $10000$ grains to ensure the accuracy of our observations.
Because the contact dynamics algorithm induces some amount of numerical
noise, the motion of a collection of grains driven
at different shear rates is not expected to reproduce exactly 
the same phase space trajectory.
In Figure~\ref{scaledledata} we show raw data of the normalized pressure $p \dot\gamma^{-2}$
as function of shear strain (strain rate multiplied by time), at a packing fraction of $0.8$ 
with no friction and at two different values of the shear rate.

According to the invariance in Newton's equations
$p \dot\gamma^{-2}$ should be independent of $\dot\gamma$, and this behavior is confirmed 
by the measurements in Figure~\ref{scaledledata}.  
Although the shear rates in the two plots differ by a factor of $10^4$, the normalized pressure is virtually identical for both systems.  Interestingly, not only do the steady state values show no shear rate dependence, but the initial transient is virtually identical for both values of $\dot\gamma$. 
The invariance in Newton's equations also predicts that $\sigma \dot\gamma^{-2}$ and $T \dot\gamma^{-2}$ are independent of $\dot\gamma$.  For all of the simulations we have carried out these predictions from the invariance are upheld-- although numerical noise often disrupts the perfect invariance for large values of shear strain, we see no change in the steady state values of normalized pressure, shear stress, or granular temperature as the shear rate is varied at constant density.  These tests ensure that the simulations we conduct respect the
invariance in Newton's equations for hard spheres.  

\begin{figure}
\psfrag{tl}{\LARGE{$\mathbf{\sigma/p}$}}
\psfrag{ml}{\LARGE{$\mathbf{\dot\gamma/\sqrt{T}}$}}
\psfrag{bl}{\LARGE{$\mathbf{T/p}$}}
\resizebox{!}{.38\textwidth}{{\includegraphics{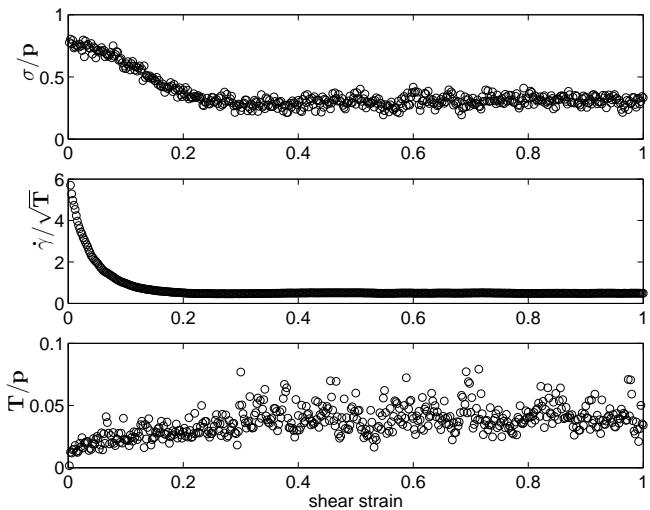}}}
\caption{\label{soverpandeoversqrtTvsshear} Invariant quantities $\sigma/p$ (top), $\dot{\gamma}/\sqrt{T}$ (middle), and $T/p$ (bottom) as a function of shear strain for a frictionless granular material at packing fraction $0.8$.}
\end{figure}

For a granular material characterized by its pressure $p$, shear stress $\sigma$, temperature $T$,
and strain rate $\dot\gamma$, we can construct three independent 
invariant quantities: $\sigma/p$, $\dot\gamma / \sqrt{T}$, and $T/p$.
In Figure~\ref{soverpandeoversqrtTvsshear} we show values of these three independent
invariant quantities as a function of shear strain for a frictionless granular material at packing fraction of $0.8$.
For all quantities, steady flow is reached by
a shear of approximately $0.5$, and we will subsequently provide 
stationary data by time-averaging our measurements between strains of $2$ and $10$.  The values of $\sigma/p$ and $T/p$ fluctuate much more than $\dot\gamma/\sqrt{T}$.  This is due to the fact that $\sigma$ and $p$ depend on the forces between grains, which are highly fluctuating in the hard sphere limit.  In our simulations with $10000$ grains we observe that the fluctuations decrease while the average value remains constant.  This suggests that in the limit of large system size, the fluctuations would disappear.  

\subsubsection{Liquid-solid transition}
\begin{figure}
\resizebox{!}{.38\textwidth}{{\includegraphics{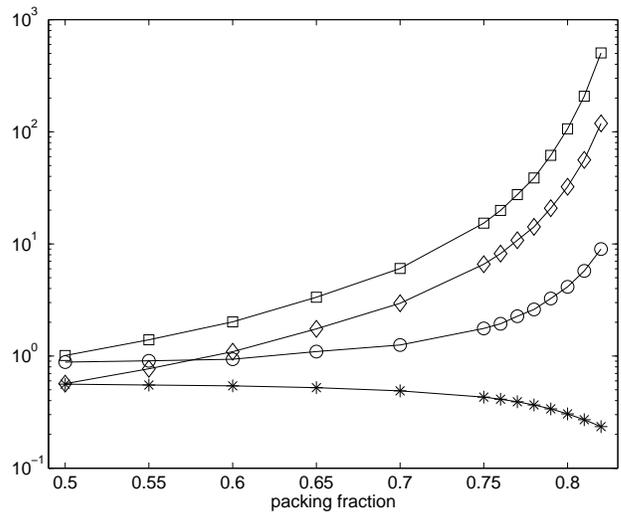}}}
\caption{\label{scaledlevspacking} Steady state values of $p \dot{\gamma}^{-2}$ (squares), $\sigma \dot{\gamma}^{-2}$ (diamonds), $T \dot{\gamma}^{-2}$ (circles), and $\sigma/p$ (stars) as a function of packing fraction for a frictionless granular material.}
\end{figure}

In Figure~\ref{scaledlevspacking} we present the steady state values of 
$p \dot{\gamma}^{-2}$, $\sigma \dot{\gamma}^{-2}$, $T \dot{\gamma}^{-2}$, and $\sigma/p$ in simple shear for a range of high packing 
fraction systems that we have studied, at zero friction.  Although there is relatively little change in these quantities for
small packing fraction, for packing fractions larger that $0.75$ there is a large increase in the values of the 
stresses and granular temperature.  Additionally, the functional form of the stresses and granular temperature changes from approximately exponential to a function that grows faster than an exponential at $\nu \approx 0.75$.

\begin{figure}
\resizebox{!}{.38\textwidth}{{\includegraphics{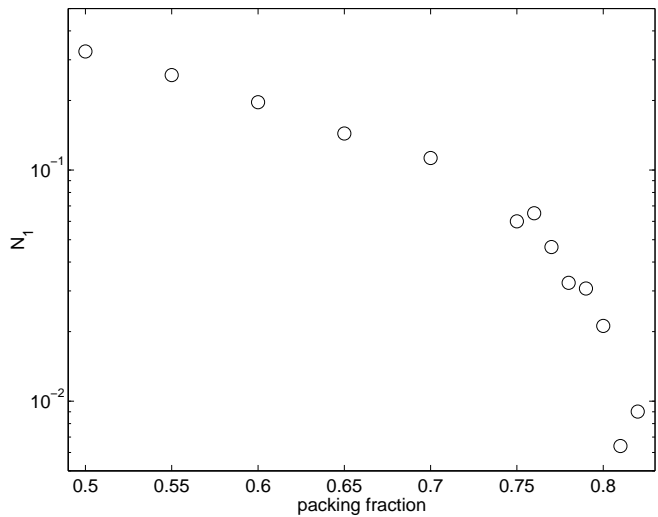}}}
\caption{\label{firstnormalstress} First normal stress difference $\mathit{N}_1$ as a function of packing fraction for a frictionless granular material.  $\mathit{N}_1>0$ for all packing fractions suggests that the granular materials are amorphous, and the decay of $\mathit{N}_1$ as a function of packing fraction suggests that non-kinetic effects dominate at high packing fraction.}
\end{figure}

Alam and Luding~\cite{AL03} have measured the steady state values 
of the first normal stress difference $\mathit{N}_1$, defined via equation~(\ref{stressbasis}),
in simple shear flow using a monodisperse collection of grains.
In dilute flows, a non-vanishing value for $\mathit{N}_1$ 
results from collisional terms and the anisotropy 
in the distribution of velocities at Burnett order~\cite{GS96}.
Alam and Luding reported that $\mathit{N}_1$ becomes negative 
at the onset of crystallization. 
We have thus measured the first normal stress difference
in our system in order to ensure
the absence of crystallization and 
as a signature of the breakdown of kinetic effects.
The observation in Figure~\ref{firstnormalstress} that $\mathit{N}_1>0$ for all packing fractions in our system
is consistent with the observation that our system remains amorphous.
The decay of $\mathit{N}_1$ over all packing fractions 
is consistent with the idea that kinetic effects become less important as
the packing fraction increases, especially after $\nu \approx 0.75$ 
when $\mathit{N}_1$ begins to quickly decay.

In our numerical simulations, we expect the collisional contributions
to the stress tensor to be negligible. A detailed study of the crossover
between collisional and non-collisional regimes will be the topic of a future
work~\cite{LoisLC05b}. For now we rely on the observation that the 
kinetic effects, as probed by the first normal stress difference, decay close to jamming.
We also refer to prior works which support the same conclusion~\cite{CCH95,PC96}.

\subsection{Flow down an inclined plane}
In the simple shear cell the density and strain rate were specified, 
while stresses and granular temperature were measured.
We now focus on flow down an inclined plane which
provides a complementary situation.
For incline flow, stresses are specified by the choice of an angle of inclination
and by the gravitational field.  Then the
profiles of velocity and velocity fluctuations are measured and grant
access to profiles of strain rate and granular temperature.

\begin{figure}
\psfrag{tl}{\LARGE{$\mathbf{\nu}$}}
\psfrag{ml}{\LARGE{$\mathbf{T}$}}
\psfrag{bl}{\LARGE{$\mathbf{\dot\gamma/\sqrt{T}}$}}
\resizebox{!}{.38\textwidth}{{\includegraphics{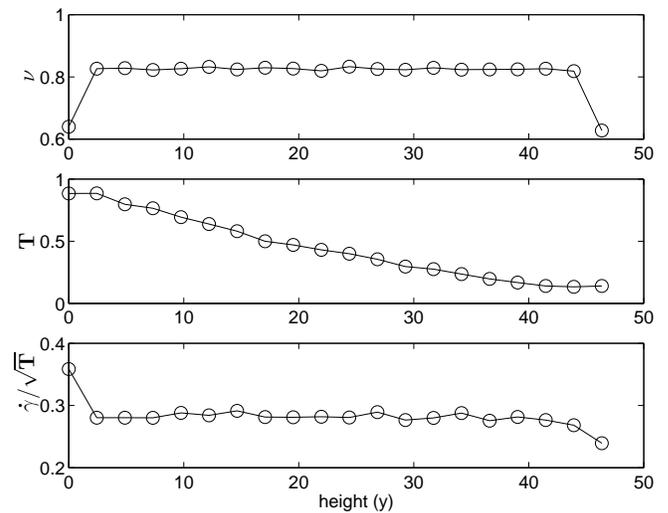}}}
\caption{\label{inclineprofs} Profiles of packing fraction (top), granular temperature (middle), and $\dot{\gamma}/\sqrt{T}$ (bottom) as a function of the height ($y$) in the pile, measured in grain diameters, for a non-frictional granular material at a $12^{\circ}$ incline.}
\end{figure}

We report in Figure~\ref{inclineprofs} the packing fraction, granular temperature, and $\dot{\gamma}/\sqrt{T}$
as a function of height for the steady flow of a non-frictional granular material at an angle of $12^{\circ}$, 
with a total height of approximately $50$ grains.  We have conducted simulations with heights ranging between $25$ and $100$ grain diameters to ensure that our results do not depend on the size of the system.

We observe that, in the bulk central region of the flow, the packing fraction profile is uniform,  
the granular temperature is linear, and $\dot{\gamma}/\sqrt{T}$ is constant.  These observations hold in our simulations
for all angles where the granular flow reaches a steady state.  We only use data from these steady state flows in this paper.
These observations are consistent with previous observations 
by Silbert et al~\cite{silbert01,silbert03}.

\begin{figure}
\psfrag{tl}{\LARGE{$\mathbf{\sigma/p}$}}
\psfrag{ml}{\LARGE{$\mathbf{\dot\gamma/\sqrt{T}}$}}
\psfrag{bl}{\LARGE{$\mathbf{T/p}$}}
\resizebox{!}{.38\textwidth}{{\includegraphics{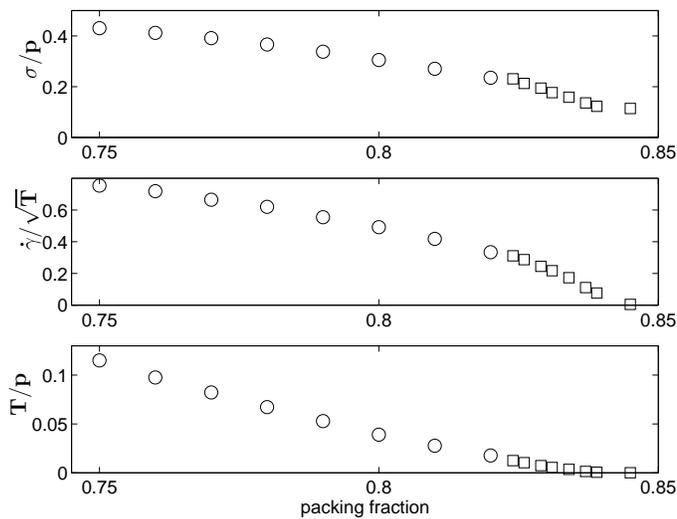}}}
\caption{ \label{matchincdensity} The values of $\sigma/p$ (top), $\dot{\gamma}/\sqrt{T}$ (middle), and $T/p$ (bottom) plotted as a function of packing fraction.  Data from simple shear flow (circles) and flow down an incline (squares) match on the same curves.  This suggests that there is a local rheology that is independent of the particular shearing geometry.}
\end{figure}

There are four quantities of interest for incline flows--  $p$, $\sigma$, $\dot{\gamma}$, and $T$-- and these lead to three independent invariant quantities $\dot{\gamma}/\sqrt{T}$, $\sigma/p$, and $T/p$.  We observe in our simulations that all of these invariants are constant in the bulk of the incline flow.  Therefore it is legitimate to compare these constant values with the constant values obtained from simple shear simulations.    
In Figure~\ref{matchincdensity} we present how the constant values of $\sigma/p$, $T/p$, and $\dot{\gamma}/\sqrt{T}$ in the bulk of the flow depend on packing fraction and compare with our results from the simple shear cell.  
The fact that data from different shear flows fall on the same curves is remarkable and suggests that one theory should be able to describe simple shear and bulk incline granular flow.

Interestingly, the data from different flows do not overlap over a large interval of packing fraction:  the flow down an inclined plane provides values at higher values of packing fraction than the simple shear cell.  This is due to (i) steady flows down an incline plane are more easily reached for lower inclinations, hence higher densities; and (ii) the simple shear deformation is more difficult to integrate numerically at higher densities, because the periodic cell induces additional constraints that the Contact Dynamics algorithm manages with difficulty.  Nevertheless, our use of two different configurations grants access to a broad range of $\dot{\gamma}/\sqrt{T}$, $\sigma/p$, and $T/p$; and the sets of data are consistent with the existence of a unique, local relation between them as apparent in Figure~\ref{matchincdensity}.

\subsection{Origin of a Local Rheology}
The excellent agreement between the two sets of data challenges 
the belief that non-local effects 
arise in dense granular flow.  The data in Figure~\ref{matchincdensity}
supports a very conservative opinion:
the motion of the grains decorrelates beyond some finite length scale, 
in accord with the fast decay of the pair correlation function (see Figure~\ref{paircorr}).

As a consequence, in the bulk of the flow down an incline, it is possible to view 
layers of granular materials as effective simple shear cells.
Such a layer of granular material at height $y$ 
responds essentially as if it was confined in a simple shear cell, in the 
absence of body forces, with sustained external stresses $\sigma(y)$ and $p(y)$.
Of course the invariance in Newton's equations, which holds exactly for
the simple shear cell, is slightly broken by the gravitational force field.
However, deep in the bulk of the flow, large confining stresses
eventually dominate over the gravitational field.
This approximate invariance suffices to predict that Bagnold's scaling
must hold for the bulk regions of incline flows
and explains the numerical data of Silbert and coworkers~\cite{silbert01,silbert03}.

Hydrodynamic equations are expected
to arise when the gradients of macroscopic variables become small
compared to the macroscopic variables themselves.
What is surprising, however, is that the locality of the rheology
is observable for the moderate heights that we can access in
our numerical simulations. 
We believe that the reason why the hydrodynamic regime seems to be observed
at accessible scales is the following:
(i) the relevance of hydrodynamics limit depends on how a macroscopic
quantity $\Psi$ compares with $\mathit{\ell} \nabla \Psi$, where
$\ell$ is a length-scale; (ii) $\ell$ is not the size of the grain
or a cluster of grains-- it is the size of the region
sampled by a given grain per strain unit, which is on 
the order of the mean-free path.
Since for a dense granular material $\mathit{\ell}/\langle R \rangle<<1$,
the validity of the hydrodynamic limit should hold over moderate heights.

\begin{figure}
\psfrag{yl}{\LARGE{$\mathbf{\sigma/p}$}}
\psfrag{xl}{\LARGE{$\mathbf{\dot\gamma/\sqrt{T}}$}}
\resizebox{!}{.38\textwidth}{{\includegraphics{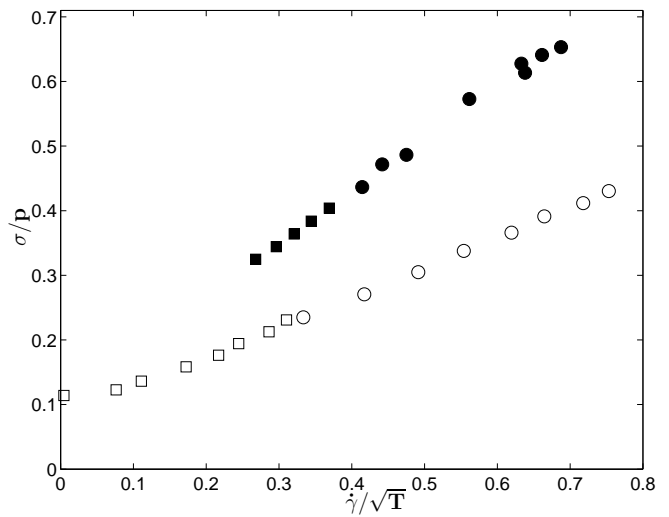}}}
\caption{\label{lemoney} $\sigma/p$ plotted against $\dot{\gamma}/\sqrt{T}$ in steady state simple shear flow (circles) and incline flow (squares).  Filled symbols correspond to frictional granular materials with coefficient of friction $\mu=0.4$ and open symbols correspond to non-frictional granular materials with $\mu=0$.}
\end{figure}

As we will see, our theory provides constitutive equations which relate 
the static friction coefficient $\sigma/p$ to the ratio $\dot{\gamma}/\sqrt{T}$.
Anticipating the following sections, we present in Figure~\ref{lemoney}
a plot of $\sigma/p$ versus $\dot{\gamma}/\sqrt{T}$, with data from both the simple shear and incline flow geometries. 
As we will argue, such a plot is expected to be the granular counterpart of a stress 
versus strain rate plot for a glassy material~\cite{lemaitre02a}, with the shear stress normalized by the pressure and the strain rate normalized by the granular temperature.
We see here that the analogy is striking: when rescaled properly the 
granular material presents typical features of normal yield 
stress fluids~\cite{larson99}.
For large values of normalized strain rate, the normalized stress is proportional to the normalized strain rate.  For small values of the normalized strain rate, the linear relationship no longer holds and there is a yield (normalized) stress at zero (normalized) strain rate.     

\section{V. Constitutive equations}

Several interesting results stem from the preceding observations:
(i) for a given
density, only invariant quantities are relevant to describe the state
of a granular flow;
(ii) when invariant quantities are considered, the granular material
displays a ``normal'' rheology of a yield stress liquid; and
(iii) the rheology measured in the Lees-Edward cell extrapolates to
the rheology measured from bulk data of incline flow.

These observations foster our hope to construct local constitutive equations
for dense granular materials. 
Before addressing the specificities of the STZ formulation 
of constitutive equations, we wish to argue that the scaling invariance 
of Newton's equations calls for a parallel scaling form of the constitutive equations.

\subsection{Scaling form for constitutive equations}
In general, constitutive equations should provide a relation between
stresses and strain-rate, and a set of state variables 
$\{\psi_i\}$ which characterize the internal structure of the granular material.
For granular materials the granular temperature $T$ plays a quite specific role as a state variable.
Other state variables are expected to account only for geometric 
properties of the granular packing.

\subsubsection{Conservation of energy}
Ogawa~\cite{ogawa78} was the first to recognize that in granular materials
the temperature cannot be prescribed by a thermal bath, but
is set by energy balance. 
Kinetic theory provides estimates 
for the energy dissipation rate in dilute systems,
hence providing approximations for the equation of motion 
which governs granular temperature.
As a result, the notion of granular temperature has been tied to kinetic theory. 

Granular temperature, however, is not reserved for the description of dilute flows.
It is relevant in all inertial flows since, as discussed previously,
there is no quasi-static limit for hard-spheres, even infinitely close to jamming.
In particular the dense flows of our simulations do not verify the assumptions 
of kinetic theory, yet in these inertial regimes kinetic energy is a perfectly relevant
physical observable and is set by energy balance.

At any time, the variation of kinetic energy results from
the balance between external work done on the system and dissipative mechanisms.
Because the system explores phase-space trajectories at a velocity proportional to $\sqrt{T}$,
the rate of energy dissipation should scale as $\sqrt{T}\,T$.  The square root
gives the frequency of dissipative events and each event dissipates an energy
proportional to $T$.  Energy is introduced to the system via the external forcing with a rate of $\sigma \dot\gamma$.
Therefore, the energy balance equation should be of the form
\begin{equation}
\phi \dot{T} = \sigma \dot{\gamma} - \alpha(\phi,\{\psi_i\}) 
T\:\frac{\sqrt{T}}{\langle R \rangle},
\quad
\label{tempeqn}
\end{equation}
with $\phi$ being the mass density.
The factor $\alpha$ is a geometric factor which depends
on the relative positions of grains, but should not incorporate
any further dependence on $T$ or the stress tensor. 
In particular, it should depend on the mass density $\phi$ and possibly on other state
variables $\{\psi_i\}$ which characterize the geometrical structure of the granular packing.
By scaling invariance, the steady state value of $\alpha$ should depend of $\phi$ only.

This equation has a similar form to the energy conservation equation derived 
in kinetic theory \cite{GD99}. This is no surprise since kinetic theory
must uphold the invariance of Newton's equations. Kinetic theory
provides an estimate for $\alpha(\phi)$ and it is possible that even 
in dense flows this estimate remains reasonable for non-frictional grains.
However, multibody collisions may induce departure of this relation 
from the predictions of kinetic theory.

\begin{figure}
\resizebox{!}{.38\textwidth}{{\includegraphics{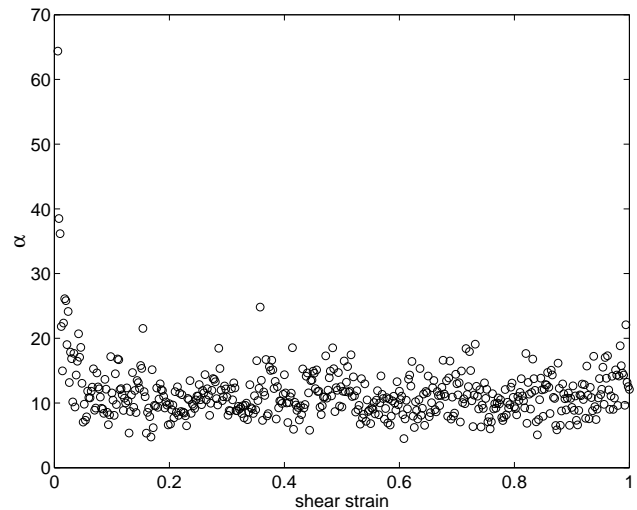}}}
\caption{\label{alphavsshear} $\alpha$ from equation~(\ref{tempeqn}) as a function of shear strain for a simple shear flow of a non-frictional granular material at packing fraction $0.8$.  For small values of shear $\alpha$ varies slightly, but quickly becomes constant at a shear strain of approximately $0.04$.}
\end{figure}

In Figure~\ref{alphavsshear} we present a plot of $\alpha$ as a function of shear strain for a non-frictional granular material in simple shear flow, calculated via equation~(\ref{tempeqn}).  We observe a dynamic in $\alpha$ at the start of the simulation which quickly disappears.  The dynamic is likely due to the effects of an unknown state variable $\psi_i$.

\subsubsection{Constitutive relation}

Since granular temperature is defined as the frequency of elementary
events along phase-space trajectories, the strain rate should scale as
\begin{equation}
\frac{\dot{\gamma}}{\sqrt{T}} = f\left( \frac{\sigma}{p},\frac{T}{p},\{\psi_i\} \right),   
\label{fconst}
\end{equation}
where $f$ denotes an unknown function and, once again, the state variables 
$\{\psi_i\}$ are purely geometric.

If we combine equation~(\ref{fconst}) with equation~(\ref{tempeqn}) 
in steady state where $\dot{T}=0$, then we see that
\begin{equation}
\sigma = \left[\frac{\alpha}{\langle R \rangle} F\left(\frac{\sigma}{p},\psi_i \right)^{-3}\right] \dot{\gamma}^2   
\label{bagnoldcouette}
\end{equation}
where $F$ is equal to the function $f$, with $T/p$ evaluated using equation~(\ref{tempeqn}).
This equation implies that if constitutive relations are written in the form of equations~(\ref{tempeqn})~and~(\ref{fconst}), then 
Bagnold's scaling is upheld.

In the rest of this section we study the STZ formulation of constitutive
equations and apply it to dense granular materials.  We will see that the STZ theory, 
when applied to granular materials, makes a prediction for the function $f$ in equation~(\ref{fconst}). 

\subsection{STZ theory}

\subsubsection{Basics}

The Shear Transformation Zone (STZ) Theory of amorphous solids was
proposed in~\cite{falk98,falk99,falk00,langer01,langer03,langer03a}
to account for the behavior of dense amorphous materials at low temperature.  
The theory is motivated by observations from 
simulations~\cite{KobayashiMT80,MaedaT81,srolovitzME81,takeuchi87,DengAY89}
and experiments~\cite{argon79b} which suggest that plastic deformation in amorphous materials
results from non-affine rearrangements of small clusters of particles~\cite{argon83}.
Additionally, Falk and Langer were able to show that there exist different 
types of zones which present a preferential
response to different orientations of shear forces. 
They introduced the densities of these zones as state variables to 
characterize the internal structure of the molecular packing.

Central to the theory is the assumption that once an STZ undergoes an elementary
rearrangement in a given direction it is unlikely that it can shear again in 
the same direction, although it can easily shear in the reverse direction.
Thus zones appear as two state systems, the states corresponding to the  
zone orientation being aligned (denoted $-$) or anti-aligned (denoted $+$) with the shear stress. 
A rearrangement corresponds
to a transition of a zone from a $\pm$-state into a $\mp$-state and vice-versa.
The plastic shear rate is given by the rate at which STZs respond to external
stresses:
\begin{equation}
\dot{\gamma} \propto R_{+} n_{+} - R_{-} n_{-},    
\label{firststz}
\end{equation}
where $R_{\pm}$ are the stress-dependent probabilities that zones of $\pm$ types
are transformed into one another.
 
This constitutive relation must be complemented with an equation of motion
for the densities $n_\pm$, which is postulated to be of the form:
\begin{equation}
\dot{n}_{\pm} = R_{\mp} n_{\mp} - R_{\pm} n_{\pm} +w (a-b n_{\pm}).   
\label{secondstz}
\end{equation}
The first two terms correspond to the transformation of STZs into their 
two possible states. The last term accounts for the fact that STZs are renewed
by the overall macroscopic deformation: it contains a creation and destruction rate,
both proportional to the plastic work $w$ of external forces per time unit.

\subsubsection{Observation of directional response}

Before applying STZ theory to granular materials, we check that 
the same qualitative observations as in~\cite{falk98}
can be performed in these systems.  Namely that non-affine motion occurs in localized regions
and that the positions of the localized regions depends sensitively on the orientation of the shear.  
The first observation motivates the choice of density of STZs as a state variable, and the second 
observation shows that each STZ has an orientation and therefore only responds to a certain orientation
of shear stress.

These observations are based on a measure of the non-affinity
of the deformation of a cluster of a few molecules or grains.  Following~\cite{falk98},
the grains undergoing non-affine
rearrangement can be determined by calculating, for each grain, the local strain rate at 
time $t-\Delta t$.  Then, by measuring the difference $D$ between the actual position at a later time $t$ and the position
predicted from the local strain rate at time $t-\Delta t$, we can determine which grains have moved non-affinely in the time period $\Delta t$.  In practice, $D$ is determined by minimizing

\begin{eqnarray}
\tilde{D}^2(t,\Delta t) = \sum_n \sum_i \Big( r_n^i(t)-r_0^i(t)-\sum_j(\delta_{ij}+\dot{\gamma}_{ij}) \nonumber \\
         \times  \big[r_n^j(t-\Delta t)-r_0^j(t-\Delta t) \big] \Big) ^2    
\end{eqnarray}
with respect to the shear rate $\dot{\gamma}_{ij}$, where the indices $i$ and $j$ are spatial coordinates and the index $n$ runs over all grains within two diameters of the reference grain, labeled by the index $n=0$.  The minimum value of $\tilde{D}$, denoted $D$, is an approximation of the local deviation from affine displacement for the reference grain in the time interval $[t-\Delta t,t]$.
If there is no non-affine motion, then the motion of each individual grain 
should be completely determined by a local shear rate and $D=0$.  
If there is non-affine motion then $D>0$.    

\begin{figure}[htbp]
\mbox{
\subfigure[]{\scalebox{.25}{\includegraphics{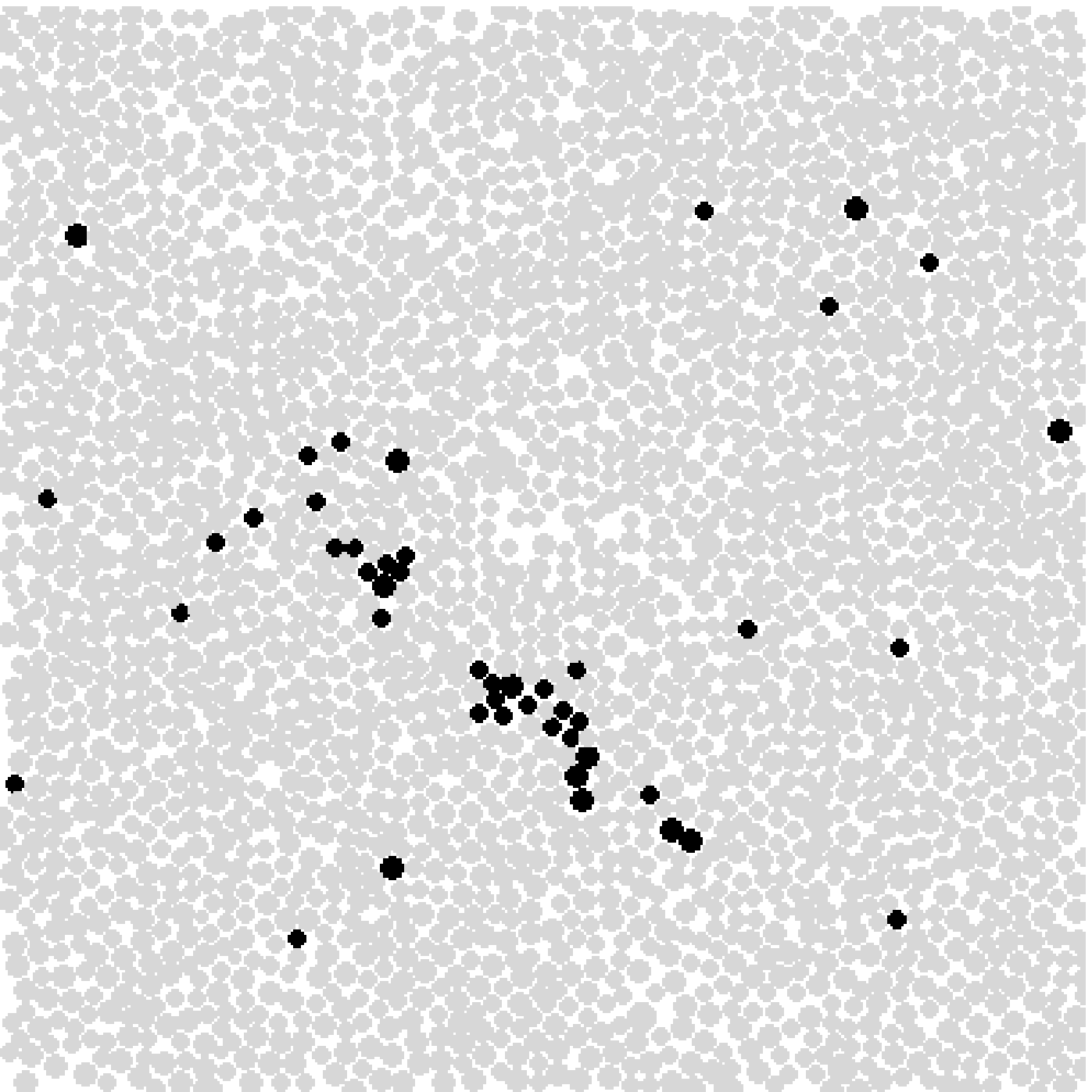}}}
\quad \quad 
\subfigure[]{\scalebox{.25}{\includegraphics{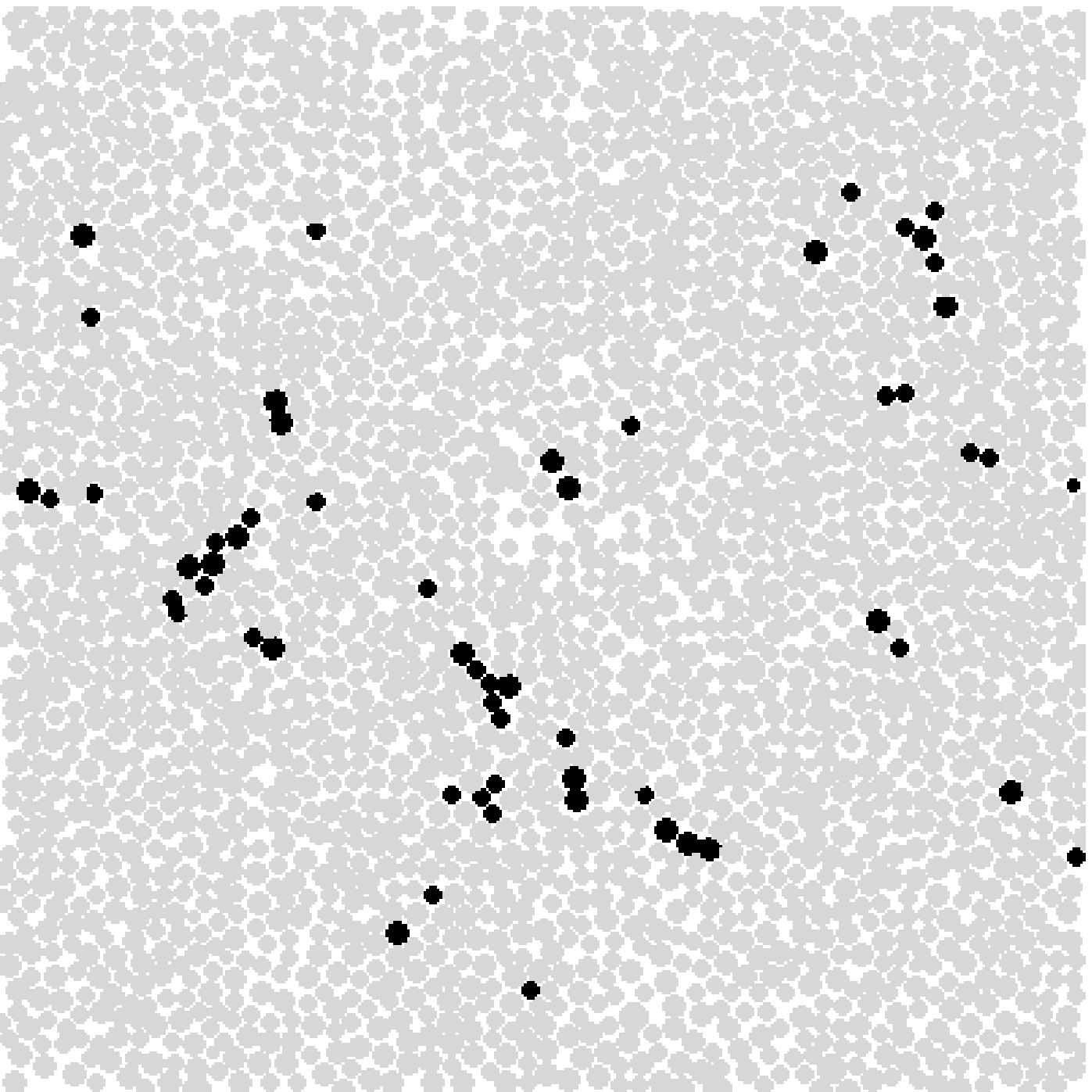}}}
}
\mbox{
\subfigure[]{\scalebox{.25}{\includegraphics{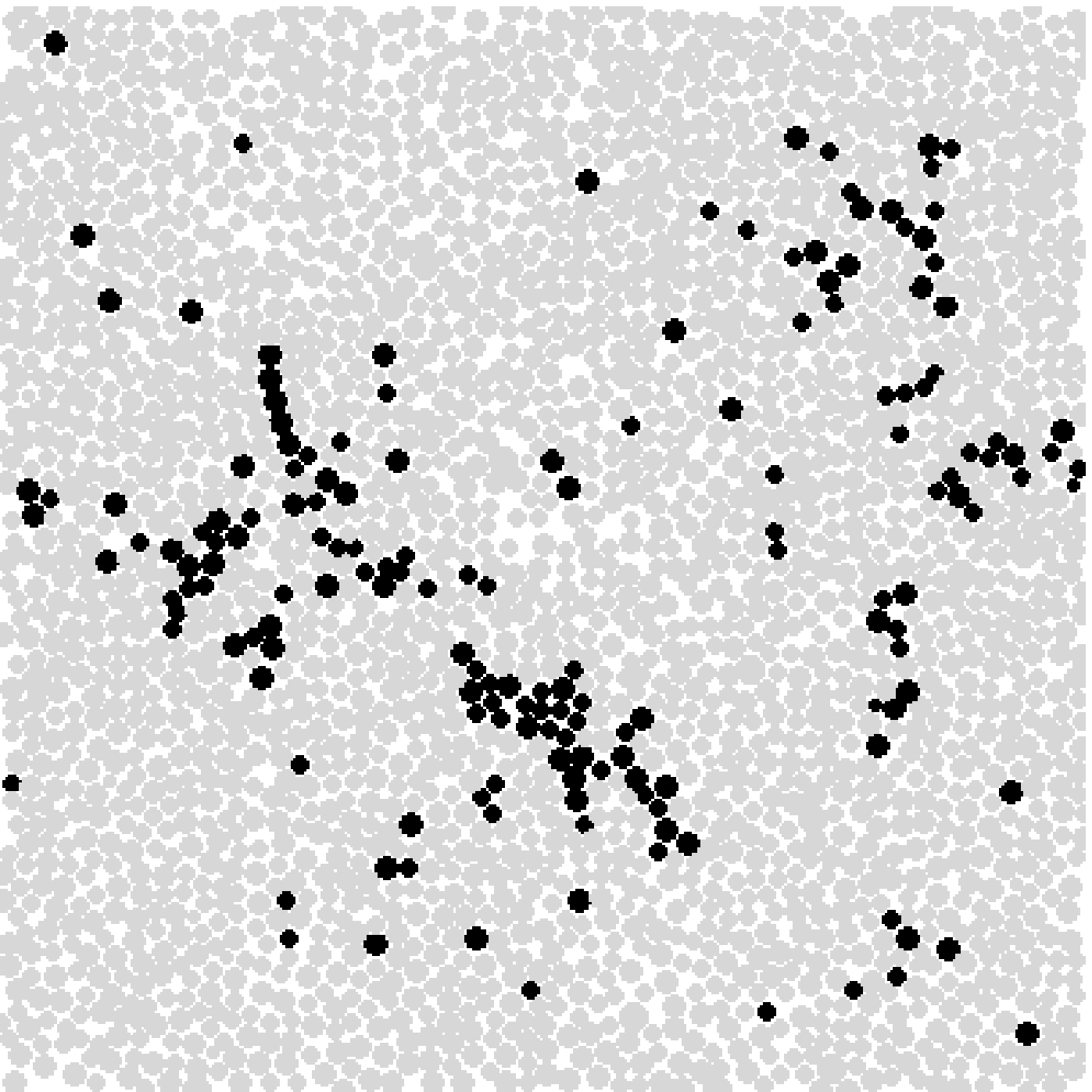}}}
\quad \quad 
\subfigure[]{\scalebox{.25}{\includegraphics{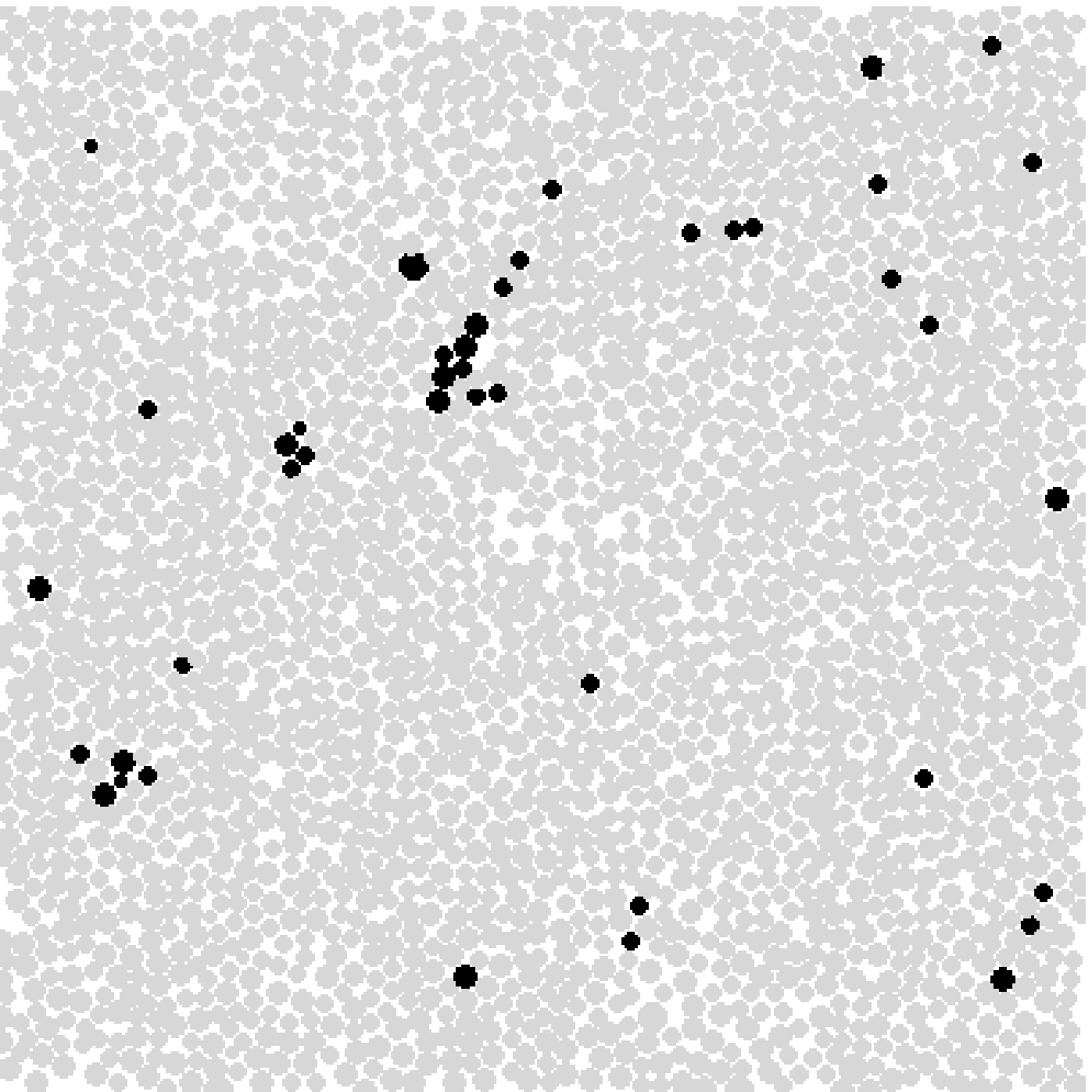}}}
}
\caption{ \label{stzfs} Screenshots of granular materials in steady state simple shear flow with grains undergoing non-affine displacement colored black.  In (a) the system is sheared from $0$ to $0.05$, in (b) the system is sheared from $.1$ to $.15$, in (c) the system is sheared from $0$ to $.15$, and in (d) the system is sheared in the opposite direction, from $0$ to $-.05$.}
\end{figure}

We have applied this test for non-affine motion to granular materials in simple shear 
to produce Figure~\ref{stzfs}.  This figure is the counterpart of
Figure~3 in~\cite{falk99} and Figure~7 in~\cite{falk98}, which were created from simulations of an amorphous Lennard-Jones solid.
Each picture has been created by shearing an identical initial arrangement
of particles in a certain direction.  $D$ is the local measure of non-affinity
obtained by comparison between the initial and final states.
If $D$ is larger than a reference value, 
the particle is said to have moved non-affinely and colored black.

In Figure~\ref{stzfs} part (a) the system is sheared from strains of $0$ to $0.05$, 
and in (b) the system is sheared to from strains of $0.1$ to $0.15$.
We notice that there is a tendency for the regions of non-affine displacement to form clusters, and the size of the resulting non-affine regions 
is about the same in both (a) and (b).
In (c) the system is sheared from strains of $0$ to $0.15$ 
and now the size of the non-affine regions increases, suggesting many more fundamental 
rearrangements of STZs in the larger time period.   
In (d) the system is sheared from strains of $0$ to $-0.05$ (in the {\it opposite} direction), starting from the 
same initial configuration as in (a).  Once again we observe that the non-affine regions
tend to form clusters.  However, in comparing (a) and (d), we notice that the
size of the non-affine regions is about the same in the two figures, but the locations are different.
If the regions undergoing non-affine displacement did not have an orientation 
we would expect non-affine motion to occur in the same location, 
regardless of the orientation of the stress.  However this is not the case 
and the data in Figure~\ref{stzfs} suggests that the regions 
that move non-affinely have an orientation.   
This is qualitative evidence that the core assumptions of STZ theory 
are upheld in granular materials.

\subsubsection{STZ Theory for Granular Materials}

The STZ densities $n_\pm$ account for structural properties
of a molecular or granular packing. They are thus expected to depend 
on the positions of grains, orientations and distribution of forces, 
orientations of velocities, but not on the overall amplitude 
of the forces or amplitude of velocities.
Following~\cite{lemaitre02a}, these functions are determined using the invariance in Newton's equations.
Since $w$ represents the plastic work done on the system per unit time 
it should be proportional to $\sigma \dot{\gamma}$.
In order to make $w$ invariant, we normalize by pressure so that 
$w=\sigma \dot{\gamma}/p$.
As for $R_{\pm}$, because of the invariance in Newton's equations 
we can separate the rate at which an STZ attempts to rearrange
from the probability that an attempt leads to a successful rearrangement.  
The attempt rate must be proportional to $\sqrt{T}$ which sets the microscopic 
event rate and the probability to rearrange is written 
as an exponential activation factor of the invariant form $e^{\pm \kappa \sigma/p}$.
This yields $R_{\pm} \propto \sqrt{T}e^{\pm \kappa \sigma/p}$.

Combining the expressions for $R_{\pm}$ and $w$ 
with equations~(\ref{firststz}) and (\ref{secondstz}), 
while making a change of variables from $n_{\pm}$ to 
$\Delta \propto n_{-} - n_{+}$ and $\Lambda \propto n_{-} + n_{+}$, 
yields the following STZ equations for granular materials:
\begin{equation}
\begin{split}
\dot{\gamma} &\propto \sqrt{T}\left(
\Lambda \sinh{(\kappa \sigma/p)}-\Delta \cosh{(\kappa\sigma/p)} \right)  \\
\dot{\Delta} &\propto {\dot{\gamma}}\left(
1-\zeta\,\frac{\sigma}{p}\Delta  \right)\\
\dot{\Lambda}&\propto \dot{\gamma}
\frac{\sigma}{p}(1-\Lambda).  
 \label{STZequations}
\end{split}
\end{equation}
$\Lambda$ denotes the total density of zones and
$\Delta$ measures the mismatch between zones of different orientations
and therefore is related to the anisotropy of the granular packing.
$\kappa$ and $\zeta$ are constants that do not depend on 
the macroscopic variables $\dot{\gamma}$, $T$, $\sigma$, or $p$.
However we would expect these constants to depend on properties of 
the grains such as shape, distributions of radii, restitution coefficients, 
friction coefficients, or other local static variables, including density.

\subsubsection{Steady states}
These equations present two types of steady
state solutions~\cite{falk98,falk00,lemaitre02a}.
One branch of solutions represents a jammed state 
$\dot{\gamma}=0$, and occurs when $\Delta/\Lambda = \tanh{(\kappa \sigma/p)}$.
The other branch of solution represents the steady flow and 
occurs when $\Lambda=1$ and $\Delta=p/(\gamma\sigma)$.

An elementary analysis of the phase diagram of this dynamical system indicates that
the jammed state is stable if and only if 
\begin{equation}
\zeta\frac{\sigma}{p}\,\tanh\left({\kappa\frac{\sigma}{p}}\right)\leq0
\end{equation} 
and the flowing state is stable otherwise.
The limit of stability occurs at a critical angle $\theta^*$, which is the solution
of
\begin{equation}
\zeta\tan\theta^*\,\tanh{(\kappa\tan\theta^*)}=0.
\end{equation}
$\theta^*$ is identified as the repose angle of our granular material.

In the steady flowing regime $\dot{\gamma} \geq 0$,
STZ theory yields the following relation:
\begin{equation}
\label{stzconstitutive}
\frac{\dot{\gamma}}{\sqrt{T}} \propto \left(\sinh{\left( \kappa \sigma/p \right)}-\frac{p}{\sigma \zeta} \cosh{\left( \kappa \sigma/p \right)} \right) 
\end{equation}
In particular, in the limit when the ratio $\frac{\dot{\gamma}}{\sqrt{T}}$ vanishes, 
$\sigma/p$ converges towards $\tan\theta^*$.
Therefore the STZ theory accommodates cases where there is a residual pressure and 
shear stress at zero shear rate, and predicts that in this case the shear stress 
will be proportional to the pressure.

\subsection{Numerical tests}

The constitutive equations for granular materials introduced in this paper are entirely specified by 
equations~(\ref{tempeqn}) and~(\ref{STZequations}).
We now compare the steady state relations predicted by these equations with our
numerical data, which provides independent access to 
$\sigma$, $p$, $T$, and $\dot{\gamma}$ in different shearing geometries.

\begin{figure}
\resizebox{!}{.38\textwidth}{{\includegraphics{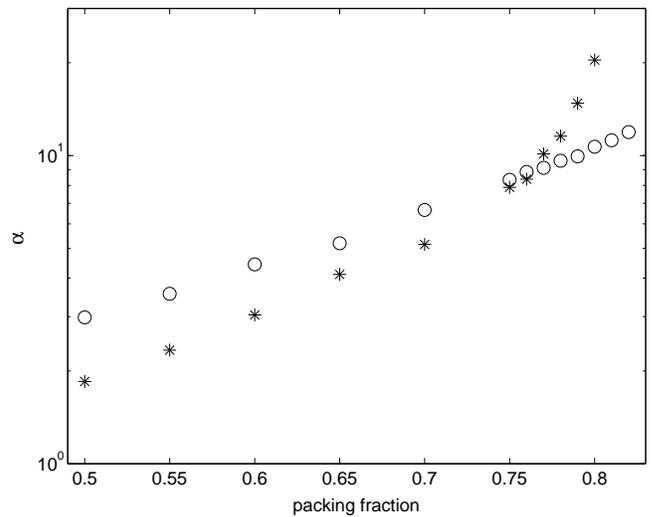}}}
\caption{\label{alphavspacking} Density dependence of $\alpha$, defined through equation~(\ref{tempeqn}), for frictional (stars) and non-frictional (circles) simple shear flows.}
\end{figure}

We first test equation~(\ref{tempeqn}) in simple shear flow.  We find that for all densities investigated, 
much like Figure~\ref{alphavsshear}, $\alpha$ is constant as a function of shear strain.
Additionally we find that the steady state value only depends on the density and friction coefficient.
We present the measured steady state values of $\alpha$ 
in Figure~\ref{alphavspacking} as a function of packing fraction,   
for frictionless ($\mu=0$) and frictional ($\mu=0.4$) granular materials.  For frictionless 
granular materials $\alpha$ appears to vary exponentially as a function of packing fraction, whereas for frictional 
granular materials $\alpha$ does not take on an obvious functional form.

\begin{figure}
\psfrag{yl}{\LARGE{$\mathbf{\sigma/p}$}}
\psfrag{xl}{\LARGE{$\mathbf{\dot\gamma/\sqrt{T}}$}}
\resizebox{!}{.38\textwidth}{{\includegraphics{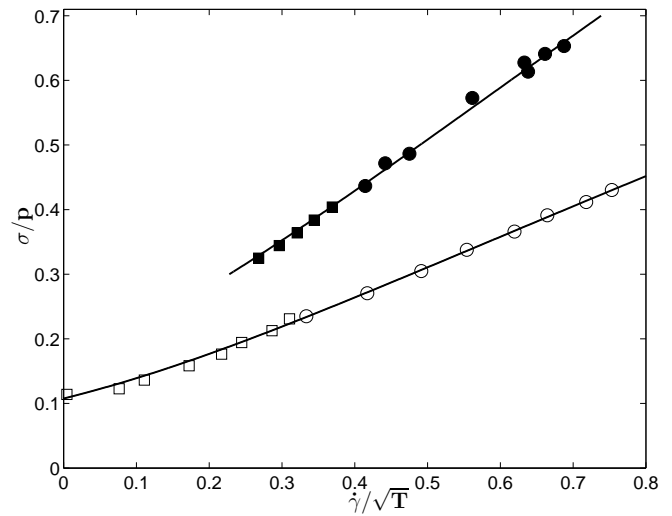}}}
\caption{\label{money} $\sigma/p$ plotted against $\dot{\gamma}/\sqrt{T}$ for frictional (filled symbols) and non-frictional (open symbols) granular materials.  The circles correspond to data from simple shear flow and the squares to data from steady state incline flow.  The line is a best fit to the steady state STZ prediction from equation~(\ref{stzconstitutive}), where only the data from the simple shear flow geometry was used to construct the best fit.  This shows that STZ theory properly predicts the outcome of incline flow experiments once the parameters (which depend only on grain properties) are determined from simple shear experiments.}
\end{figure}

Next, we test the steady state STZ prediction from equation~(\ref{stzconstitutive}). 
In Figure~\ref{money} we have plotted numerical data of $\sigma/p$ as a function of $\dot\gamma/\sqrt{T}$ for frictional and non-frictional 
granular materials in both simple shear flow and incline flow.
The line drawn through the data is a fit 
to equation~(\ref{stzconstitutive}).    
The fit matches the data from both flowing geometries very well . 

To construct the fit, we have used the data from {\it the simple shear cell only}.
This fit permits us to extrapolate the rheology for the flow down an incline plane,
even at the approach of the repose angle.  This shows that an accurate determination of the coefficients from the STZ equation~(\ref{stzconstitutive}) in one shearing geometry allows for prediction in a different shearing geometry.    

\section{VI. Conclusion}

We have implemented numerical simulations of dense granular flows
in order to clarify the microscopic origin of jamming and the specific assumptions
needed to construct the STZ formulation of constitutive equations for dense
granular flows.

Our first goal was to obtain detailed information on dense granular flows,
in particular on the locality of the constitutive relation.
For this purpose we implemented two different simulation configurations:
a simple shear cell with Lees-Edwards periodic boundary conditions and SLLOD
equations of motion, and the flow of a granular material down a rough incline.
In both configurations we measured the components of the stress tensor,
packing fraction, strain rate, and granular temperature.

We have also observed the normal stress difference and shown
that it vanishes in dense flows so that, in this regime, shear stress and pressure 
completely characterize the stress tensor.
In each configuration we studied, the rheology of dense granular materials 
is entirely specified by five quantities: shear stress $\sigma$, pressure $p$, 
packing fraction $\nu$, strain rate $\dot\gamma$ and granular temperature $T$.
In the shear cell $\nu$ and $\dot\gamma$ are prescribed,
and in the flow down an incline
$\sigma$ and $p$ are prescribed.
The rheology of the granular material requires us to specify three relations 
which determine the remaining unknowns.
There are many equivalent ways these relations can be formulated,
but we wished to formulate them in a way that emphasized
the invariance of Newton's equations and the expected scaling 
form of constitutive equations.

We found, as anticipated in~\cite{lemaitre02a}, that the relation between quantities
$\dot\gamma/\sqrt{T}$ and $\sigma/p$ is remarkably similar to the stress-strain rate
relation in a normal elasto-plastic transition.
We were surprised to observe the quality of the fit to the data of Figure~\ref{money}
with expression~(\ref{stzconstitutive}) without any density dependence of the parameters of the theory.
We would have expected that these parameters might involve strong density dependence.
This observation strengthens the hypothesis that the distinguishing feature
of granular materials is the way energy balance is prescribed and 
otherwise, once the variables are appropriately rescaled, they behave like ``normal'' materials.

Another relation emerges naturally from the expectation that energy balance in granular
materials should take the form in equation~(\ref{tempeqn}).
The density-dependent parameter $\alpha$ accounts for the rate of dissipation of energy and
we would expect it to be reasonably captured by collisional integrals 
of the type used in kinetic theory (at least for non-frictional materials).
A proper derivation of energy dissipation in dense flows now appears as an accessible
yet challenging issue that we wish to examine in future works.

Finally, a third relation remains to be specified. This relation can be, for example,
a prescription for $\sigma/p$ or $\dot\gamma/\sqrt{T}$ as a function of packing fraction. 
We observed this relation in Figure~\ref{matchincdensity}, but did not attempt to model it.
This missing relation should arise from some understanding of how the rheology depends
on the local density. Such a property might arise from studies of free-volume formulations
of the rheology of 
dense granular materials.~\cite{lemaitre02b,lemaitre02c,lemaitre03,LemaitreC04,Baz04,Baz05} 
We did not wish to elaborate this aspect in order to focus on the stress-strain rate
relation, which is the most immediate prediction of STZ theory.

As we see, the STZ theory correctly predicts one among three relations 
which are necessary for a complete description of dense granular flows. 
This relation, coupled to Bagnold's scaling, suffices to account for 
the shape of the velocity profile down an incline. 
However it does not account for the overall amplitude 
of the velocity profile or the uniform value of the density 
as a function of $\tan\theta$.

Note that we have compared here data from a Lees-Edward cell with
the rheology of a flow down an inclined plane in the limit where the 
height of the granular material is large. 
The agreement between the two sets of data indicates that there
is no long-range structure which governs the flow. This does not mean, however, that
no gradient or diffusive terms should ever appear in the final form of 
the rheology of granular flows. In particular, we expect that a diffusive term
should arise in equation~(\ref{tempeqn}), but in the large height limit 
this diffusive term does not appear in the bulk rheology.

This brings us to the observations by Pouliquen~\cite{Pou99} 
of a relation $h_{\rm stop}(\theta)$ which governs jamming.
These observations have led to the idea that non-local constitutive equations 
might be required to capture the rheology of granular materials.
However, we expect these observations to be ``normal'' finite size effects.
For example, it was observed in~\cite{lemaitre02a} that an exponentially 
decaying kernel in place of $w$ in equation~(\ref{secondstz}) was sufficient to account 
for large values of $h_{\rm stop}(\theta)$ when $\theta\to\theta^*$.
The analytical form of the kernel used in this work was consistent with an exponential
decay of correlation in the system, again consistent with the notion that short-range 
correlations suffice to understand these effects.
But other theories also successfully introduce finite length scales to account for macroscopic
phenomena. For example, Bazant's 'spot' model~\cite{Baz04,Baz05}
relates such a length scale to the size of diffusive regions of high free-volume.
Such concepts are very interesting as they provide routes to understanding non-trivial
phenomena in granular physics, relying on very plausible forms of finite size 
corrections to hydrodynamics.

Another puzzling observation by Pouliquen was that the amplitude
of the velocity profile seemed to scale as $1/h_{\rm stop}(\theta)$.
As we saw previously, this amplitude is governed by $\alpha(\nu)$ and a missing
relation between {\it e.g.\/} $\sigma/p$ and $\nu$. We thus cannot provide
an interpretation for these observations.

Our observations on a model system strongly suggest that the definition 
of a local rheology for granular materials is, in principle, possible.
Moreover, we see emerging from our analysis some state variables and their equations
of motion. This now opens a very exciting route toward a set of local 
hydrodynamic equations for dense granular materials: such equations 
will offer a predictive tool to further address fundamental and practical
questions in numerous situations where flow equations for granular
materials are much needed.

\section{Acknowledgements}
This work was supported by the William. M. Keck Foundation, the MRSEC program of the NSF under Award No. DMR00-80034, the James. S. McDonnell Foundation, NSF Grant No. DMR-9813752, the Lucile Packard Foundation, and the Mitsubishi Corporation.


\end{document}